\documentclass[aps,amsmath,amssymb,reprint,superscriptaddress]{revtex4-1}
\usepackage{graphicx}% Include figure files
\usepackage{dcolumn}% Align table columns on decimal point
\usepackage{bm}% bold math
\usepackage[english]{babel}
\newcommand{\p}{$\%$}

\newcommand{\pn}{$\mathrm{R{_{N_2}}}$}

\newcommand{\dNbN}{$\mathrm{\delta-NbN}$}
\newcommand{\Tc}{$\mathrm{T_{C}}$}

\newcommand{\tg}{$\mathrm{t_{2g}}$}
\newcommand{\e}{$\mathrm{e_{_g}}$}
\newcommand{\vn}{$\mathrm{V_{_N}}$}
\newcommand{\vnb}{$\mathrm{V_{Nb}}$}
\newcommand{\Ni}{$\mathrm{N_i}$}
\newcommand{\Nii}{$\mathrm{N_{2i}}$}
\usepackage{tikz}
\usepackage{amsfonts,amssymb,amsmath, geometry, url}
\usepackage{dcolumn}
\usepackage{float}
\begin{document}
%\setstcolor{red}
\title{Microscopic Origin of Structural Disorder in \dNbN: Correlation of Superconductivity and Electronic Structure}
\author {Shailesh Kalal}
\affiliation{UGC-DAE Consortium for Scientific Research, University Campus, Khandwa Road, Indore-452 001, India}
\author{Sanjay Nayak}
\affiliation{Thin Film Physics Division, Department of Physics, Chemistry and Biology (IFM), Linköping University, SE-581 83, Linköping, Sweden}
\author{Akhil Tayal}
\affiliation{Deutsches Elektronen-Synchrotron DESY, Notkestrasse 85, D-22607 Hamburg, Germany}
\author{Jens Birch}
\affiliation{Thin Film Physics Division, Department of Physics, Chemistry and Biology (IFM), Linköping University, SE-581 83, Linköping, Sweden}
\author {Rajeev Rawat}
\affiliation{UGC-DAE Consortium for Scientific Research, University Campus, Khandwa Road, Indore-452 001, India}
\author {Mukul Gupta}
\email{mgupta@csr.res.in}
\affiliation{UGC-DAE Consortium for Scientific Research, University Campus, Khandwa Road, Indore-452 001, India}

\date{\today}

\begin{abstract}

Rock-salt type niobium nitride (\dNbN) is a well-known superconductor having superconducting transition temperature (\Tc) $\approx$~18\,K and a large superconducting gap $\approx$~3\,meV. The \Tc\ of \dNbN\ thin film exhibits a large scattering irrespective of the growth conditions and lattice parameter. In this work, we investigate the atomic origin of suppression of \Tc\ in \dNbN\ thin film by employing combined methods of experiments and ab-initio simulations. Sputtered \dNbN\ thin films with different disorder were analyzed through electrical resistivity and x-ray absorption spectroscopy. A strong correlation between the superconductivity and the atomic distortion induced electronic reconstruction was observed. The theoretical analysis revealed that under N-rich growth conditions, atomic and molecular N-interstitial defects assisted by cation vacancies form spontaneously and are responsible for the suppression of \Tc\ in \dNbN\ by smearing its electronic densities of states around Fermi level.
\end{abstract}

\maketitle

Superconducting niobium nitride (NbN) thin films have been extensively used to fabricate modern technological devices like: single photon detector~\cite{gol2001picosecond}, hot electron bolometer~\cite{baselmans2004doubling}, Josephson junction~\cite{yu2002coherent}, high field superconducting magnet~\cite{kampwirth1985application}, nano–electromechanical systems and high–pressure devices~\cite{blase2009superconducting} etc. The choice of NbN for several technological applications has been motivated due to its relatively higher superconducting transition temperature (\Tc~$\approx$~18\,K) and high superconducting energy gap ($\Delta(0)\approx$~3\,meV) among transition metal nitrides (TMNs)~\cite{keskar1971,kamlapure2010}. The superior mechanical stability and ease of fabrication of NbN with cost-effective technique such as sputtering is another reason for its popular choice to fabricate devices~\cite{kamlapure2010measurement}. Among its several polymorphs, \dNbN\ (space group: $\mathit{Fm\bar{3}m}$) shows the highest value of \Tc\ owing to its stronger electron-phonon coupling caused by larger electronic densities of states around Fermi energy (E$_F$) and lower Debye temperature ($\Theta_D$)~\cite{babu2019electron,zou2016discovery}.
\par
One of the crucial issues in the development of superconducting NbN based technology is to achieve the optimum \Tc. In the literature, it is well documented that the growth techniques and conditions play a vital role in determining \Tc. Polakovic et al.~\cite{polakovic2018_IBA} have reported that, \Tc\ as a function of N$_2$ gas pressure shows a dome like behavior with maximum value of $\approx$ 14\,K at a specific range of N$_2$ concentration (17-20\p). Similar results have been reported by Choudhuri et al.~\cite{chaudhuri2010infrared} where \Tc\ of NbN thin films maximizes at certain nitrogen partial pressure. Through Hall measurements Chockalingam et al.~\cite{chockalingam2008} demonstrated  that \Tc\ of NbN thin films deposited at different \pn\ was governed by carrier density caused by Nb and/or N vacancies concentration. However, this work does not shed light on a drastic reduction of the carrier density with samples grown at higher N$_2$ partial pressure (\pn). Similar results have been widely reported in the literature~\cite{dane2017bias,shiino2010improvement,chand2009temperature}. Often, the atomic disorder has been attributed to the reduction of the \Tc\ in this superconductor~\cite{chand2009temperature,chockalingam2009evolution,carbillet2020}.
\par
Thus, to uncover the microscopic origin of the widely speculated structural disorder and consequently its effect on superconductivity, we synthesize \dNbN\ thin films using a dc-magnetron sputtering (details of the deposition parameters are discussed in section I of supplementary material (SM)~\cite{SI}). Disorder in the films is tuned by varying the \pn\ during the deposition process. Samples have been thoroughly characterized using complementary characterization tools. A combined approach of experiments and first-principles simulations is adopted to reach the conclusion.

\begin{table} [!ht] \centering
	\caption{\label{tab:1} Measured lattice parameter (LP) and superconducting transition temperature (\Tc) of \dNbN\ samples. The \Tc\ could not be observed within the instrumental limit for the sample grown at nitrogen partial pressure (\pn) = 30, 65, 100\p.}\vspace{2mm}
	\begin{tabular}{lll}
		\hline\hline
		\pn&LP&\Tc\\
		(\%)&({$\pm$~0.006~\AA})&(K)\\
		\hline
		16&4.376&12.8\\
		25&4.415&6.9\\
		30&4.421&-\\
		65&4.486&-\\
		100&4.505&-\\
		Theoretical (this work)&4.41&--\\
		\hline
		Experimental~\cite{keskar1971}&4.40&17.3\\
		Theoretical~\cite{babu2019}&4.45&18.2\\			
		\hline\hline
	\end{tabular}
\end{table}

\begin{figure} \center \vspace{-1mm}
	\includegraphics [width=0.4\textwidth]{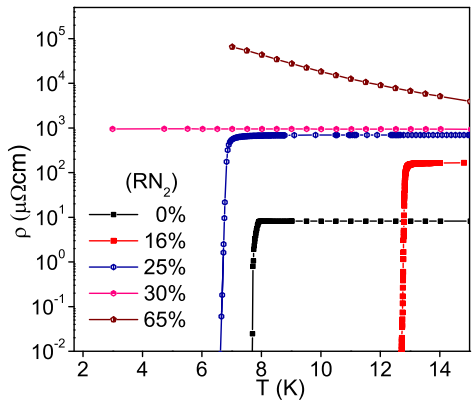}
	\caption{\label{fig:RT} Temperature dependent electrical resistivity ($\rho$) of samples deposited at \pn~=~0, 16, 25, 30, 65\p. Here, \pn~=~100\p\ sample
    not included as its $\rho>$10$^5$.}\vspace{-1mm}
\end{figure}

\begin{figure*} [!ht]
	\center \vspace{-1mm}
	\includegraphics [width=0.8\textwidth]{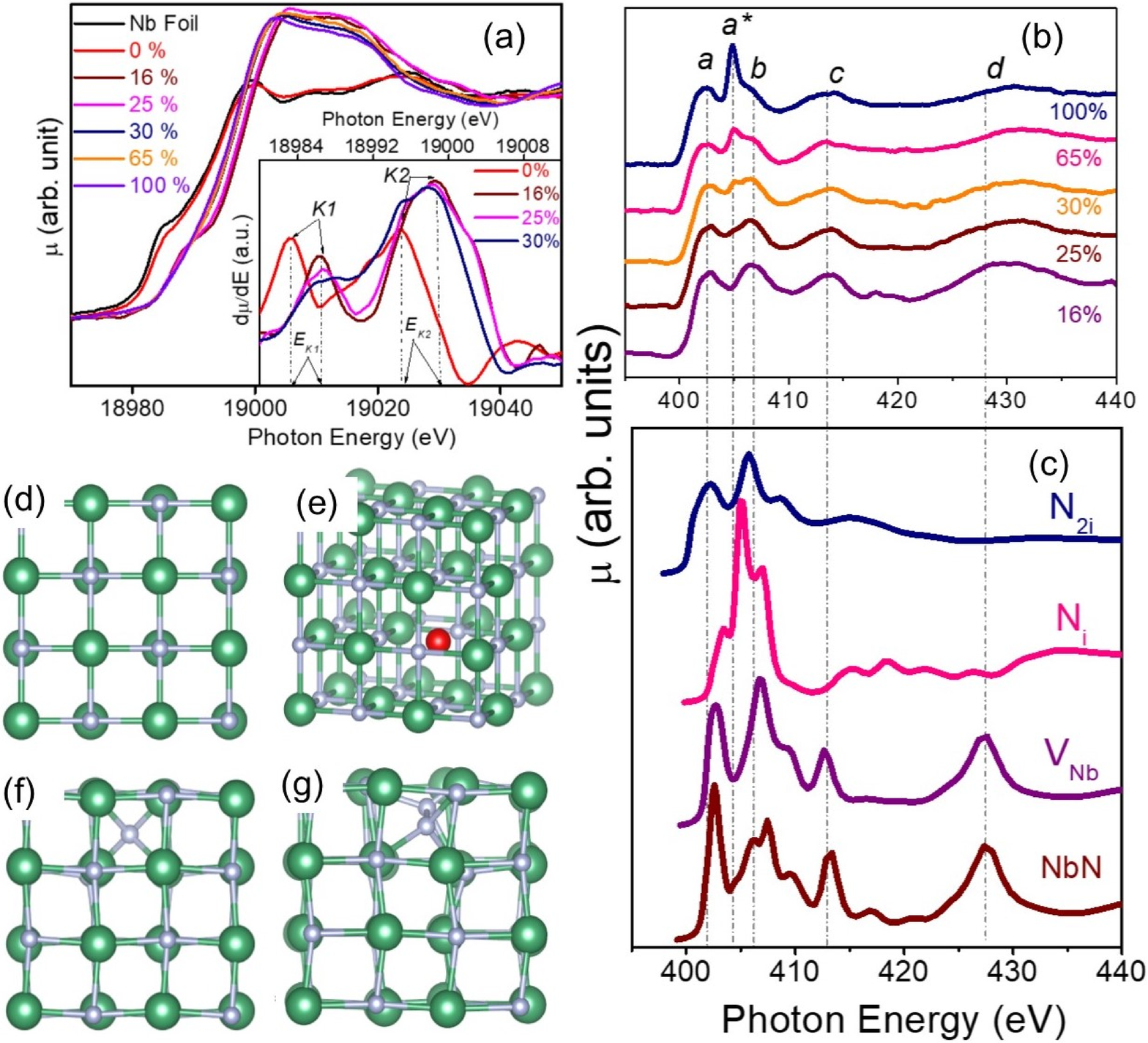}
	\caption{\label{fig:XAS} (a) Nb K-edge XANES spectra with inset showing first derivative. (b) N K-edge XANES spectra and (c) simulated N K-edge XANES spectra for various defect configurations. Ball-and-stick models of different \dNbN\ configurations used in DFT and multiple scattering theory calculations are presented: (d) pristine \dNbN, (e) with Nb vacancy (\vnb), (f) with interstitial nitrogen (\Ni) and (g) with molecular nitrogen (\Nii).}\vspace{-1mm}
\end{figure*}

A detailed report on structural and transport characterization of grown films has been published elsewhere~\cite{Kalal2020} where XRD result confirms that the sample grown at \pn~=~0\p\ is a pure Nb with bcc structure and the sample grown at \pn~=~16, 25, 30, 65, 100\p\ have a single phase of NbN with rock-salt type crystal structure (\dNbN). The \Tc\ (defined as the point where $\rho$ falls down to 10\p\ of its normal value) of the grown films are obtained from standard four probe temperature dependent electrical resistivity ($\rho$) measurements are presented in Fig.~\ref{fig:RT}. The values of \pn, LP and \Tc\ are given in table~\ref{tab:1} where a strong correlation between structure and superconductivity is noticed. Here with increase in \pn, the LP is continuously increasing and consequently a reduction in the \Tc\ is seen.
\par
In order to understand the effect of growth conditions on the electronic structure, we performed element specific x-ray absorption near edge spectroscopy (XANES) measurements both at Nb and N K-edges as shown in Fig.~\ref{fig:XAS} (a) and (b), respectively. We note that Nb K-edge splits into two components $\textit{K1}$ and $\textit{K2}$ (see inset of Fig.~\ref{fig:XAS} (a)). The initial rise $\textit{K1}$ is due to the transitions from 1$\textit{s}$ core level to unoccupied admixed 4$\textit{d}$-5$\textit{p}$ levels while the second absorption rise $\textit{K2}$ arises due to  transitions from the 1$\textit{s}$ core level to the Laporte-allowed states of pure 5$\textit{p}$ symmetry~\cite{muller_XAS,NbN_EXAFS}. The absorption edge of sample deposit at \pn~=~16\p\ is shifted to higher energy by 3~\,eV as compared to elemental Nb (see inset of Fig.~\ref{fig:XAS} (a)), which indicates that Nb atom bears positive charge due to the formation of NbN compound. Further, with \pn~$\geqslant$25\p\ $\textit{d}$ band is delocalizing around the Fermi level ($\textit{E$_{K1}$}$). 
\par
Similarly, we observed clear changes in the line shape of N K-edge spectra with increase in \pn (see Fig.~\ref{fig:XAS} (b)). A sharp transition at threshold of around 400~\,eV arises for the sample deposit at \pn~=~16\p. Here, we note presence of electronic sub-band transition levels labeled as $\textit{a}$, $\textit{b}$, $\textit{c}$, $\textit{d}$. DOS analysis reveals (see Fig.~S2 of SM) these features are results of transition from N-1$\textit{s}$ core level to unoccupied N-2$\textit{p}$ orbitals. It is well-known that the TMNs in octahedral bonding coordination (e.g. \dNbN) metal $\textit{d}$ orbitals splits into two electronic sub-bands (\e\ and \tg) owing to its hybridized characteristics of N-2$\textit{p}$ orbitals~\cite{chen1997}. Feature $\textit{a}$ (centred at 402.8~\,eV) in N K-edge XANES spectra is a consequence of hybridized N-2$\textit{p}$ and \tg\ level of Nb-4$\textit{d}$ orbitals, feature $\textit{b}$ (centered at 406.8~\,eV) is a consequence of hybridized N-2$\textit{p}$ and \e\ level of Nb-4$\textit{d}$ orbitals. The feature $\textit{c}$ and $\textit{d}$ arises due to higher order hybridization between N-2$\textit{p}$ and Nb-5$\textit{s}$-5$\textit{p}$ orbitals. Further at \pn~=~25\p, feature $\textit{b}$ become broad and at \pn~=~30\p, sharp $\textit{a*}$ feature (404.9~\,eV) arises in between $\textit{a}$ and $\textit{b}$, whose intensity is gradually increases with increase in \pn.
\par
To identify the atomic origin of above mentioned features, we simulate N K-edge XANES spectra for multiple defect configurations using multi scattering theory (see Fig.~S1 of SM~\cite{SI}). We obtained the relaxed atomic structure of various point defects: (i) isolated N-vacancy (\vn), (ii) multiple N-vacancy (2\vn), (iii) Nb  vacancy (\vnb), (iv) N interstitial (\Ni), (v) N antisite ($\mathrm{N_{Nb}}$), (vi) Schottky type defect ($\mathrm{V_N - V_{Nb}}$), and (vii) interstitial N$_2$ molecules (\Nii) from SIESTA codes and used them to construct atomic cluster for simulation of ab-initio XANES spectra  (computational details of simulations are given in section II of SM~\cite{SI}). First, to establish the credibility of numerical parameters used in simulations, we obtained N K-edge spectra for pristine \dNbN. Clearly, features $\textit{a}$, $\textit{b}$, $\textit{c}$, and $\textit{d}$ are reproduced in simulations. The experimental spectrum obtained for the sample grown at \pn~=~16\p\ is in well agreement with the theoretically calculated spectra of \dNbN\ having \vnb\ (see Fig.~\ref{fig:XAS} (b) and (c)). A thorough comparison between experimental (Fig.~\ref{fig:XAS} (b)) and simulated (Fig.~\ref{fig:XAS} (c)) XANES spectra of N K-edge, revealed that appearance of $\textit{a*}$ peak centred around 404.9~\,eV for the sample grown at higher \pn\ is probably a signature of either \Ni\ or \Nii\ defects in \dNbN.

\begin{figure} [!ht]
	%	\centering \vspace{-1mm}
	\includegraphics [width=0.46\textwidth]{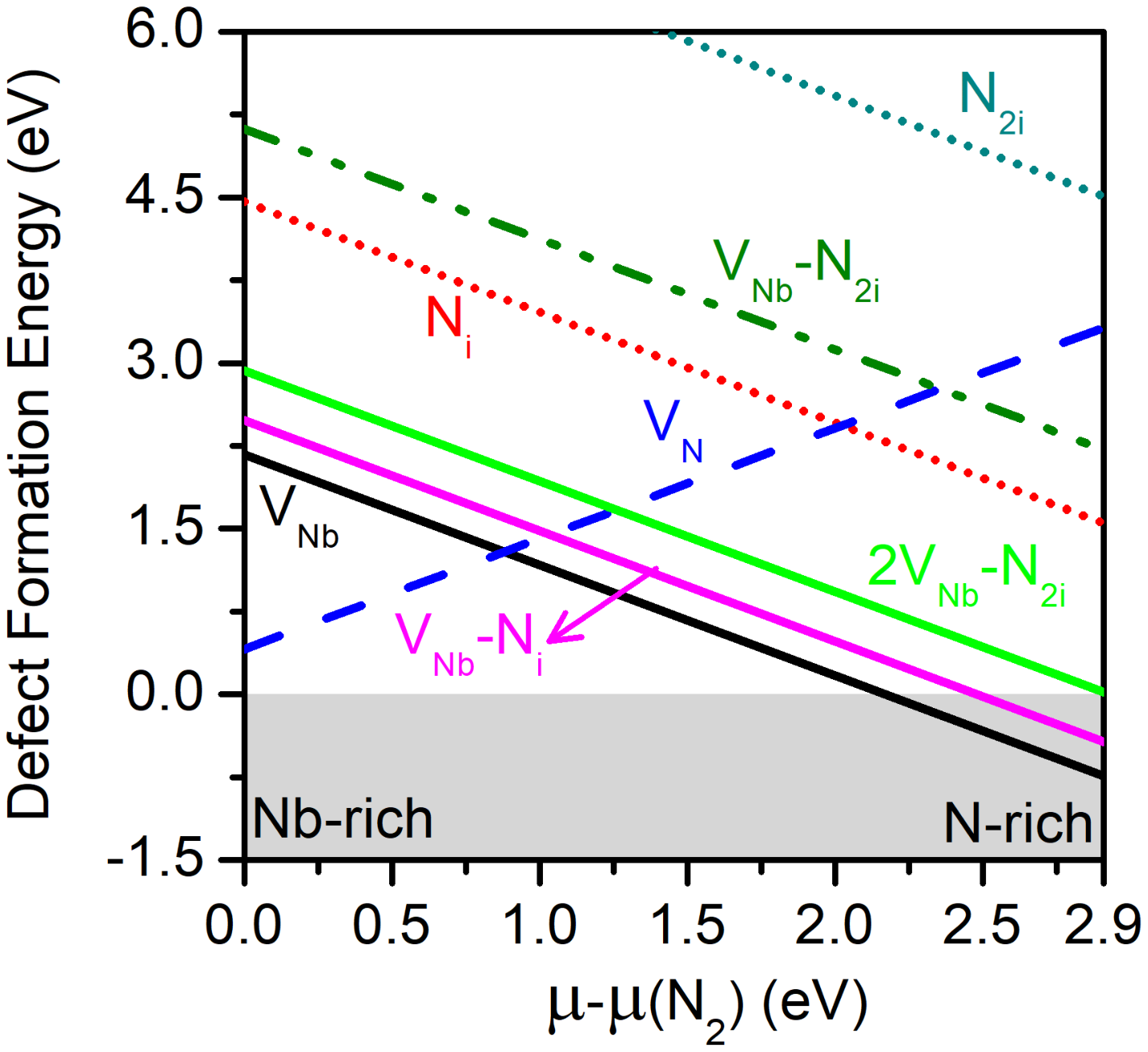}
	\caption{\label{fig:def} Formation energy as a function of chemical potential of N$_2$ for Nb vacancy (\vnb), nitrogen vacancy (\vn), interstitial nitrogen (\Ni), molecular nitrogen (\Nii), combination of niobium vacancy with interstitial nitrogen (\vnb-\Ni), combination of niobium vacancy with molecular nitrogen (\vnb-\Nii), and combination of multiple niobium vacancy with interstitial nitrogen (2\vnb-\Nii) defect configurations.}\vspace{-1mm}
\end{figure}

\begin{figure}
	%\centering \vspace{5mm}
	\includegraphics [width=0.425\textwidth]{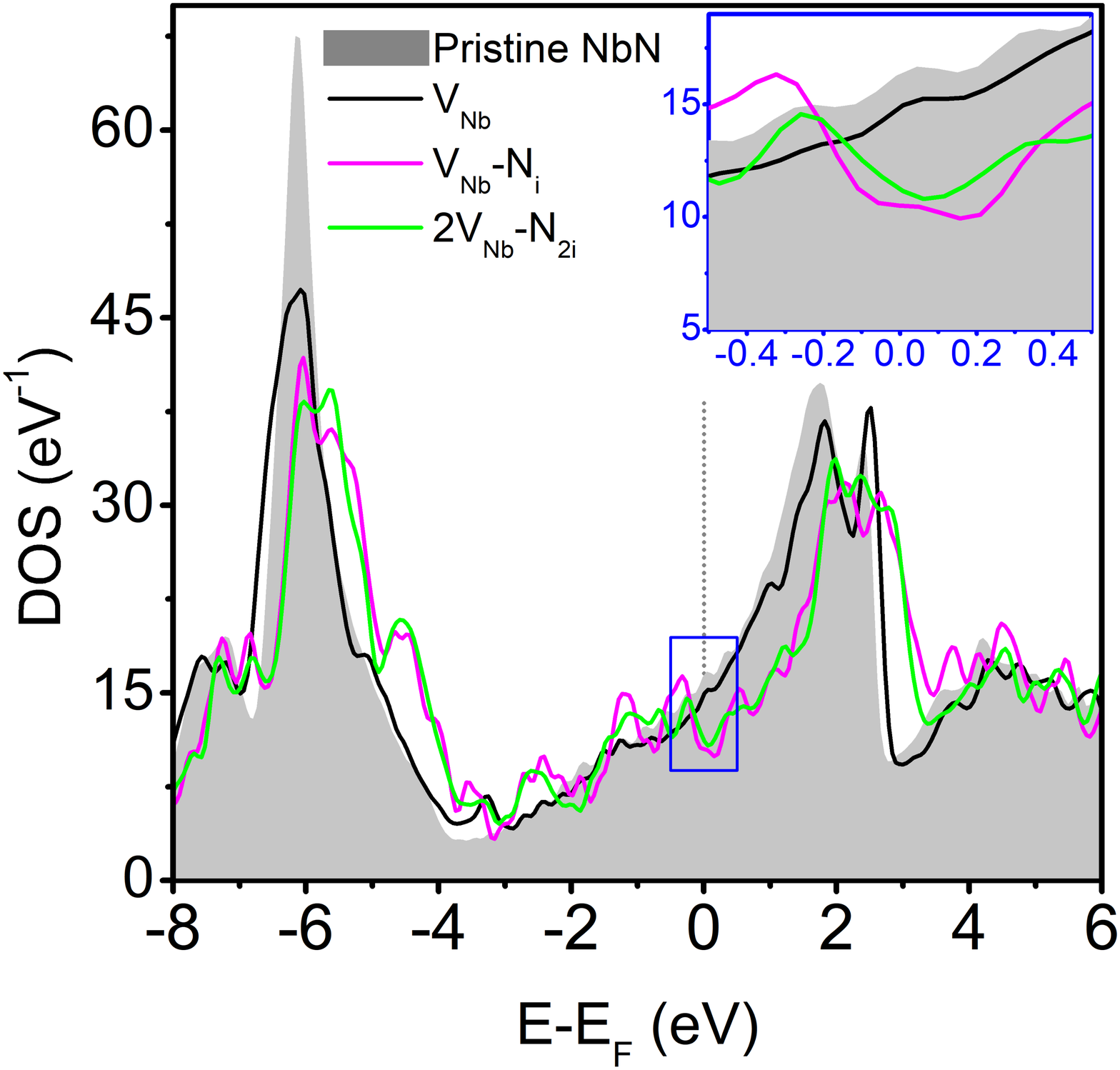}
	\caption{\label{fig:DOS} Calculated total density of state (DOS) for stoichiometric $\delta$-NbN along with Nb vacancy (\vnb), combination of Nb vacancy (\vnb) with interstitial nitrogen (\Ni), and combination of multiple Nb vacancy with molecular nitrogen ((2\vnb-\Nii) configurations. Inset shows expanded view of Fermi level. Here, Fermi level is set to 0~\,eV.}\vspace{-1mm}
\end{figure}

Gall et al.~\cite{balasubramanian2018energetics} studied the energetic of defect formation in TMNs and suggested a cation vacancy (\vnb) is the most stable defect in \dNbN, consistent with our estimation of formation energy (FE). FE plot for various defect configurations is shown in Fig.~\ref{fig:def}. Estimated FE of \vnb\ under N-rich growth condition is -0.74\,eV. Work of Gall et al.~\cite{balasubramanian2018energetics} predicted that FE of \Ni\ is very high, which is again consistent with our estimation of high FE~=~1.55\,eV at N-rich conditions. Further we find even under N-rich conditions, FE of molecular nitrogen (\Nii) in \dNbN\ is very high (4.51\,eV). High FE of \Ni\ and \Nii\ in \dNbN\ clearly suggests that they will not form spontaneously and their concentration in \dNbN\ should be negligible. Interestingly, we find that when \vnb\ and \Ni\ are in a complex form (\vnb-\Ni) their FE reduces and under N-rich condition, we estimate it to be -0.43\,eV (see Fig.~\ref{fig:def}). Also the estimated binding energy (BE) of \vnb-\Ni\ is too high (1.26\,eV). The positive BE indicates the preferential stability of the defect complex and higher the magnitude means better the stability (details of method to estimate BE is discussed in section II to SM~\cite{SI}). Similarly when \Nii\ incorporated near \vnb\ site, it forms \vnb-\Nii\ defect complex and its FE is 2.21\,eV (see Fig.\ref{fig:def}). The estimated BE of the\vnb-\Nii\ is 1.56 eV. The FE of \Nii\ further reduces to 0.02\,eV when it get coupled with two nearest neighbour \vnb\ sites (see Fig.\ref{fig:def}) with a very high BE of 5.96\,eV. These observations clearly establish that the interstitial atomic (N) and molecular nitrogen (N$_2$) can be stabilized in \dNbN\ through cation (Nb) vacancies.
\par
Further, we analyze the effect of these defects to the atomic structure of \dNbN. The obtained relaxed LP of \dNbN\ is noted to be 4.41\AA. We find that 3.125\p\ of \vnb\ in \dNbN\ reduces the unit cell volume by 0.06\p. While \vnb-\Ni\ increases the unit cell volume by 0.025\p\ and defect complex \vnb-\Nii\ causes shrinking of unit cell volume by 0.020\p. A comparison between experimentally obtained LP (and hence volume, see table~\ref{tab:1}) and theoretically computed volumes suggests that \vnb-\Ni\ is predominant in the samples grown with \pn~$>$~16\p\ and their concentration increases with increase in \pn. The smaller LP (see table~\ref{tab:1}) of \dNbN\ thin film grown at \pn~=~16\p\ suggest a presence of higher concentration of \vnb\ in it. Also the estimated lattice relaxation energy, $\mathrm{\Delta E_c}$[= $\mathrm{E_{tot}}$~(without ionic relaxation)- $\mathrm{E_{tot}}$~(with ionic relaxation)] of \dNbN\ with point defects \vnb, \vn, \Ni, \vnb-\Ni, \Nii, \vnb-\Nii, and 2\vnb-\Nii\ is 15, 10, 102, 107, 163, 160, and 172\,meV/atom, respectively. Higher values of $\mathrm{\Delta E_c}$ for interstitial defects suggest a substantial lattice distortion of \dNbN, in agreement with the increment of disorder in the \dNbN\ films with increase in \pn.
\par
Next, we shall discuss the role of these point defects (if any) on the superconducting properties of \dNbN\ thin films. Using the value of electron-phonon coupling constant ($\lambda$), \Tc\ for the strong coupling superconductors can be obtained via McMillan-Allen-Dynes formalism~\cite{allen1975superconductivity,babu2019electron,PhysRev.167.331,chockalingam2008}, given by:
\begin{equation} \label{Tc}
\mathrm{T_c = \dfrac{\omega_{log}} {1.2} exp\Big[\dfrac{-1.04(1+\lambda)}{\lambda-\mu^\ast({1+0.62\lambda})}\Big]}
\end{equation}
where $\mathrm{\omega_{log}}$ is a logarithmic average of phonon frequency, $\mu^\ast$ is the averaged screened electron-electron interaction. The $\lambda$ further calculated as $\mathrm{\lambda = [N(\epsilon_F)/ \langle \omega^2 \rangle] \sum_i \langle I^2 \rangle _i/M_i}$, where $\mathrm{M_i}$ is the atomic mass of $\mathrm{i^{th}}$ atom and $\mathrm{\langle I^2 \rangle _i}$ is the square of the electron-phonon coupling matrix element averaged over the Fermi surface~\cite{PhysRev.167.331,allen1975superconductivity}. $\mathrm{N(\epsilon_F)}$ is the electronic density of states at the Fermi level. The $\mathrm{\langle \omega^2 \rangle}$ can be further approximated as $\mathrm{0.5\Theta_D^2}$, where $\mathrm{\Theta_D}$ is the Debye temperature~\cite{liu2017first}. From eq.~\ref{Tc}, it is quite evident that \Tc\ is sensitive to the $\mathrm{N(\epsilon_F)}$. Thus, we computed the electronic DOS of NbN with previously determined dominant defects and presented in Fig.~\ref{fig:DOS}. A strong smearing in the electronic structure is visible due to the formation of \vnb-\Ni\ or 2\vnb-\Nii\ defects complex in \dNbN\ (see Fig.~\ref{fig:DOS}). DOS calculations shows that $\mathrm{N(\epsilon_F)}$ of pristine \dNbN\ ($2\times 2\times 2$ supercell) is 16.26\,states.eV$^{-1}$. For 3.125\p\ of \vnb, $\mathrm{N(\epsilon_F)}$ reduces to 14.75\,states.eV$^{-1}$. The computed $\mathrm{N(\epsilon_F)}$ for \vnb-\Ni\ and 2\vnb-\Nii\ are 10.50 and 11.15~\,states.eV$^{-1}$, respectively. We estimate \Tc\ by substituting the values of $\mathrm{\omega_{log}}$ (=269\,K), $\Theta_D$ (=637\,K), $\mu^\ast$(=0.10) computed for \dNbN\ in Ref.~\onlinecite{babu2019electron} and the normalized electronic DOS from our simulations in eq.~\ref{Tc}. For pristine \dNbN\ Gou et al. computed \Tc~=~18.26\,K, a little higher than experimentally obtained ones~\cite{babu2019electron}. Using the SIESTA, computed $\mathrm{N(\epsilon_F)}$ of \dNbN\ with 3.125\p\ \vnb\ in eq.\ref{Tc}, estimate of \Tc\ to be 15.78\,K. The experimental \Tc\ of \dNbN\ sample grown at \pn=16\p\ is 12.8\,K, hinting concentration of \vnb\ is higher than 3.125\p\ in it. Further using the $\mathrm{N(\epsilon_F)}$ of \vnb-\Ni\ and 2\vnb-\Nii\ configurations in eq.~\ref{Tc} results into \Tc\ of 7.18\,K and 8.44\,K, respectively.  These values are in excellent agreement with the experimentally obtained \Tc~=~6.9\,K of sample grown at \pn=25\p. The absence of superconducting transition (down to 3\,K) in the samples grown at \pn$\geq$30\p\ is possibly due to the presence of a large disorder in the films, which occurs in \dNbN\ films due to the presence of a large concentration of N-interstitial related defects. We estimate a 50\p\ reduction in $\mathrm{N(\epsilon_F)}$ as compared to the pristine \dNbN\ can push the \Tc\ to below 3\,K. 

\par

Thus, here we unveil the atomic structure of disordered \dNbN\ responsible for suppression of superconducting transition temperature. Under N-rich growth conditions, spontaneously formed cation vacancies are responsible for the stabilization of N-interstitial defects in \dNbN\ thin films which are otherwise unfavourable with high formation energies. The positive binding energy of the cation vacancies and anion interstitial defect complex further cements their bonding in the crystal. Formation of the N-interstitial defect complex in \dNbN\ causes strong smearing of electronic structure by creating atomic disorder in the films and thus a significant reduction in the $\mathrm{N(\epsilon_F)}$ which strongly influencing the electron-phonon coupling strength and consequently reduces the \Tc. Increase in the \Tc\ of vacuum annealed \dNbN~\cite{cukauskas1989structural,carter1987thermal,farrahi2019effect} and degradation of superconducting properties at N$_2$ atmosphere~\cite{hatano1988effects,oya2013superconducting} further support our proposed mechanism. Based on above analysis we suggest that to obtain \dNbN\ with high \Tc, films should be grown at Nb-rich conditions to avoid N interstitial defects and annealing of samples in a vacuum is recommended to eliminate residual N atoms. 
\par
In summary, we have uncovered the microscopic origin of growth parameter dependence on the \Tc\ of \dNbN. By probing the electronic structure of disordered \dNbN\ samples, we identify point defect complexes consisting of cation vacancies with atomic anion interstitial [\vnb-\Ni] and cation vacancies with interstitial molecular nitrogen [n\vnb-\Nii] are responsible for suppression of the \Tc\ in NbN films grown at higher nitrogen partial pressure (\pn). The suppression of the \Tc\ is caused by smearing of electronic structure and reduction of electronic DOS around Fermi energy due to the formation of point defect complex. We show that stabilization of atomic and molecular nitrogen in \dNbN\ is assisted by cation vacancies. Estimated \Tc\ of \dNbN\ with dominant defects identified from first-principles simulations are in good agreement with the experimentally obtained values.     
      
\section*{Acknowledgments}
We thank L. Behera, R. Sah and A. Wadikar for technical help provided in experiments. We are thankful to A. K. Sinha, Alok Banerjee, and D. M. phase for support and encouragement. This work is supported through India-DESY project.

%\bibliography{TMN}

%merlin.mbs apsrev4-1.bst 2010-07-25 4.21a (PWD, AO, DPC) hacked
%Control: key (0)
%Control: author (72) initials jnrlst
%Control: editor formatted (1) identically to author
%Control: production of article title (-1) disabled
%Control: page (0) single
%Control: year (1) truncated
%Control: production of eprint (0) enabled
%

\end{document}